\documentclass[twocolumn,aps,showpacs,floatfix,superscriptaddress]{revtex4}
\usepackage{amsmath,amssymb,wasysym,graphicx,capt-of,ifthen,calc}
\usepackage{enumerate}
\usepackage{float}
\begin{document}

\title{Voter Models on Heterogeneous Networks} 
\author{V.~Sood}
\affiliation{Complexity Science Group, University of Calgary, Calgary,
  Canada, T2N 1N4} 
\author{Tibor Antal}
\affiliation{Program for Evolutionary Dynamics, Harvard University,
Cambridge, MA~ 02138, USA} 
\author{S.~Redner} 
\affiliation{Center for Polymer
Studies and Department of Physics, Boston University, Boston, MA~ 02215, USA}

\begin{abstract}
  We study simple interacting particle systems on heterogeneous networks,
  including the {\em voter model\/} and the {\em invasion process}.  These
  are both two-state models in which in an update event an individual changes
  state to agree with a neighbor.  For the voter model, an individual
  ``imports'' its state from a randomly-chosen neighbor.  Here the average
  time $T_N$ to reach consensus for a network of $N$ nodes with an
  uncorrelated degree distribution scales as $N\mu_1^2/\mu_2$, where $\mu_k$
  is the $k^{\rm th}$ moment of the degree distribution.  Quick consensus
  thus arises on networks with broad degree distributions.  We also identify
  the conservation law that characterizes the route by which consensus is
  reached.  Parallel results are derived for the invasion process, in which
  the state of an agent is ``exported'' to a random neighbor.  We further
  generalize to biased dynamics in which one state is favored.  The
  probability for a single fitter mutant located at a node of degree $k$ to
  overspread the population---the fixation probability---is proportional to
  $k$ for the voter model and to $1/k$ for the invasion process.
\end{abstract}
\maketitle

\section{Introduction}\label{sec:intro} 

Recent studies of statistical physics models on complex networks have
elucidated the effect of heterogeneity in link structure on dynamical
properties and critical behavior.  For scale-free networks, the source of
heterogeneity is the broad distribution of node degrees, where node degree is
defined as the number of links attached to a node.  This dispersity leads,
for example, to a vanishing percolation threshold \cite{CEAH}, allows
epidemics to thrive even with a vanishingly small infection rate \cite{PV},
and causes the Ising model to be ordered at all temperatures \cite{DGM}.

For many of these models, the dynamics can be understood by accounting for
the broadness of the degree distribution within a mean-field description.  In
this article we show how to implement such an approach for two of the
simplest interacting particle systems on heterogeneous networks, namely, the
{\em voter model\/} (VM) \cite{L,K92} and its close relative the {\em invasion
  process} (IP) \cite{C05}, as well as their generalizations to biased
evolution \cite{M,nowak,WB,LHN}.

For the VM and the IP, each node of an $N$-node network can be in one of two
discrete states: $\mathbf{0}$ and $\mathbf{1}$.  In a social context, these
states represent two possible opinions of the individual at that node.  In a
biological context, the states represent the phenotype of the individual,
with state $\mathbf{0}$ representing {\em resident\/} individuals and
$\mathbf{1}$ for {\em mutant} individuals.  The evolution consists of
changing the state of a node at a rate that depends on its local environment.
The difference between the VM and IP is in the order in which the ``invader''
node (whose state is adopted) and the ``receiver'' node (whose state changes)
are chosen.  This ordering is immaterial on degree-regular graphs, such as
regular lattices, but it plays an essential role on heterogeneous degree
networks.  Understanding the basic difference between the VM and IP on such
graphs is one of the main goals of this work.

We will also study the {\em biased\/} voter model and the {\em biased\/}
invasion process in which there is a preference for one of the two states.
This generalization describes the evolution of a fitter mutant in an
otherwise homogeneous resident population.  We will determine the probability
for a single fitter mutant to overspread a population with biased VM and
biased IP dynamics.  Our main result here is that the probability for a
single fitter mutant at a node of degree $k$ to overspread a population---the
{\em fixation probability}---is proportional to $k$ for the VM and to $1/k$
for the IP.

In Sec.~\ref{sec:model}, we define the models of this paper.  In
Sec.~\ref{sec:vm-cg}, we summarize the properties of the VM on the complete
graph.  We then investigate the VM on heterogeneous networks in
Sec.~\ref{vmk}, including the complete bipartite graph, the two-clique graph,
and general degree-heterogeneous graphs.  In Sec.~\ref{sec:IP}, we treat the
complementary IP.  Finally, in Sec.~\ref{sec:evo}, we investigate the biased
versions of the VM and the IP and determine the fixation probabilities on
general heterogeneous networks.

\section{Formulation of the Models}\label{sec:model}

\subsection{Update Rules}

\noindent VM evolution consists of the following two steps: 
\begin{enumerate}[(i)]
\itemsep -1ex
\item pick a random node (a voter);
\item the voter adopts the state of a random neighbor.
\end{enumerate}
In the closely related IP the evolution steps are:
\begin{enumerate}[(i)]
\itemsep -1ex
\item pick a random node (an invader);
\item the invader exports its state to a random neighbor.
\end{enumerate}
We also mention an intermediate model, {\em link dynamics} (LD), whose
evolution steps are:
\begin{enumerate}[(i)]
\itemsep -1ex
\item pick a random link;
\item one of the nodes on the link, adopts the state of the other end node.
\end{enumerate}
In these models, the time is incremented by $1/N$ in each update.  Thus
$N$ updates corresponds to each node being updated once, on average, after
which the time increases by 1.  Steps (i) and (ii) are then repeated {\it ad
  infinitum} or until the system reaches consensus, an event that is certain
to occur in a finite time when the network is finite.

\begin{figure}[ht]
\begin{center}
\includegraphics[width=0.35\textwidth]{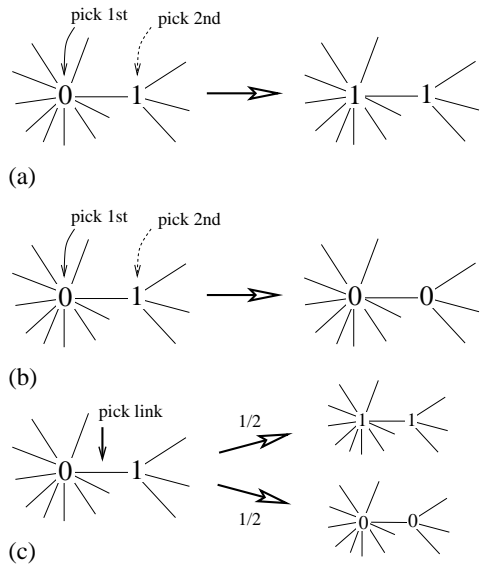}
\end{center}
\caption{Illustration of the update rules.  (a) Voter model (VM): a voter is
  chosen at random and it adopts the state of random neighbor.  (b) Invasion
  process (IP): a randomly chosen voter exports its state to a neighbor.  (c)
  Link dynamics (LD): an active link is randomly picked and a randomly chosen
  node at the end of this link is updated.}
\label{model}
\end{figure}

The interpretation of VM dynamics is that individuals lack any
self-confidence.  In an update, an individual therefore consults one of its
neighbors and adopts the neighbor's state.  On the contrary, in the IP an
individual imposes its state to one of its neighbors.  Here we can think of
the selected individual as replicating, and its offspring invades and
replaces the individual at a neighboring node.  While the different update
details of the three models might appear superficially trivial, we shall show
that these differences are fundamental when the dynamics occurs on
degree-heterogeneous networks.

In general, such a network may be specified by its adjacency matrix
$\mathbf{A}$, with $A_{x y}=1$ if nodes $x$ and $y$ are connected, and $A_{x
  y}=0$ otherwise.  The degree distribution of such a heterogeneous network
is specified by
\begin{equation}
 n_k=\frac{N_k}{N} ~,
 \label{degreedis}
\end{equation}
where $N_k$ is the number of nodes of degree $k$ and $N$ is the total number
of nodes in the network.  The moments of the degree distribution are then
\begin{equation}
 \mu_m = \frac{1}{N} \sum_x k_x^m = \sum_k k^m n_k\,.
\end{equation}
Special cases of relevance for the VM and IP are the average degree $\mu_1$,
the second moment $\mu_2$, and the average inverse degree $\mu_{-1}$.
Normalization also fixes the zeroth moment $\mu_0=1$.

Let $\eta$ represent the state of the entire network, and define $\eta(x)$,
which can take the values 0 or 1, as the state of node $x$.  In each update
event in the three models, the state of a single node changes from
$\mathbf{0}$ to $\mathbf{1}$ or vice versa (Fig.~\ref{model}).  We represent
by $\mathbf{\eta}_{x}$ the state of the system that results after changing
the state of the node at $x$:
\begin{equation}
\label{ecology:flip}
  \eta_{x}(y)= 
\begin{cases}
    \eta(y)& y \neq x\\
    1-\eta(x)& y = x
\end{cases}
\end{equation}
Transitions are specified by the probability that the state of node $x$
changes, which we term a flip event.  For the VM, IP, and LD, the transition
probability at node $x$ is
\begin{equation}
  \label{master}
  \textbf{P} [\eta \to \eta_{x}] = 
  \sum_{y} \frac{A_{x y}}{N\mathcal{Q}}\,[\Phi(x,y)+\Phi(y,x)],
\end{equation}
where $\Phi(x,y)\equiv \eta(x)[1-\eta(y)]$, and the quantity
\begin{equation*}
A_{x y}\,[\Phi(x,y)+\Phi(y,x)]
\end{equation*}
is non-zero only when nodes $x$ and $y$ are connected (the factor $A_{xy}$)
and in opposite states (expression in square brackets), so that an update can
actually occur.  Finally $\mathcal{Q}$ is
\begin{equation}
\label{cases}
 \mathcal{Q} \equiv \begin{cases}
   k_x& \textrm{~~~~~VM},\\
   \mu_1 & \textrm{~~~~~LD},\\
   k_y  & \textrm{~~~~~IP}.
 \end{cases}
\end{equation}
For the VM, the factor $(N\mathcal{Q})^{-1}$ in \eqref{master} accounts for
first choosing node $x$ with probability $1/N$, and then any of its neighbors
$y$ with probability $1/\mathcal{Q}=1/k_x$.  Conversely, in the IP, one first
chooses node $y$ (a neighbor of $x$) with probability $1/N$, and then one
chooses $x$ with probability $1/\mathcal{Q}=1/k_y$.  In LD, each link is
chosen with the same probability $2/N\mu_1$ and then one of the nodes at the
end of this link is updated with probability $1/2$, which then leads to
$\mathcal{Q}=\mu_1$.  As we shall discuss, VM, IP, and LD dynamics are
distinct for networks with heterogeneous node degrees.

The above models have been defined as discrete-time processes, where one node
is chosen in each update step.  Alternatively, these models can be formulated
as continuous-time processes by increasing the time by a random increment
that is chosen from an exponential distribution with mean $1/N$.  This
stochastic update is equivalent to each update event occurring with unit rate
in the three models.  For both discrete and continuous time dynamics, the
fixation probabilities and the average time to consensus (for long times) are
the same.  However, there is an essential difference between discrete and
continuous time dynamics for biased models that will be discussed in
Sec.~\ref{sec:evo}.

\subsection{Conservation Laws and Exit Probability}

An important aspect of the VM, IP, and LD is that each model has its own
dynamically conserved quantity.  This conservation law determines the
fundamental {\em exit probability} $\mathcal{E}_\mathbf{1}(\rho)$, namely,
the probability that a finite system with an initial density $\rho$ of
$\mathbf{1}$s reaches a consensus of all $\mathbf{1}$s.  This quantity is
also called the {\em fixation probability} in the biology literature.

The simplest case is LD, for which the state of the system changes only when
an active link---where the nodes at the ends of the link are in opposite
states---is chosen.  Because the invader and the receiver are assigned
randomly, the probability of increasing or decreasing the number of
$\mathbf{1}$ nodes (mutants) are the same.  The dynamics is thus equivalent
to a symmetric random walk on the integers, with absorbing states at $0$ and
at $N$ mutants.  Because of the symmetry of the random walk, the initial and
final densities of nodes in state $\mathbf{1}$ are the same.  Consequently,
the exit probability is
\begin{equation}
\label{martingale}
  \mathcal{E}_\mathbf{1}(\rho)= \rho=\frac{1}{N}\sum_x \eta(x) \,. 
\end{equation}
This result is also valid for the VM and the IP on degree-regular graphs, due
to their equivalence to LD when every node has the same degree.

We now extend Eq.~\eqref{martingale} to general networks.  Consider the
average change in $\eta(x)$ at node $x$, $\langle\Delta\eta(x)\rangle$.  Here
the angle brackets denote the average over all realizations of the update
dynamics.  This change equals the probability that $\eta(x)$ increases from 0
to 1 minus the probability that $\eta(x)$ decreases from 1 to 0.  Hence
\begin{equation}
\label{Deta}
\langle\Delta\eta(x)\rangle = \left[1-2\eta(x)\right] 
\mathbf{P}[\eta\rightarrow\eta_x]\,.
\end{equation}
Substituting the transition probability from Eq.~\eqref{master}, we obtain
\begin{equation}
 \langle\Delta\eta(x)\rangle = \sum_{y}\frac{A_{x y}}{N\mathcal{Q}}
\left[\eta(y)-\eta(x)\right],
\end{equation}
where we use the fact that $\eta(x)^2=\eta(x)$.  The change in the average
density in the entire network, $\langle \Delta \rho\rangle$, is obtained by
summing over all nodes $x$ to give
\begin{equation}
\label{drho-av}
\langle \Delta \rho\rangle = \sum_x \langle \Delta \eta(x)\rangle =   
\sum_{x,y} \frac{A_{x y}}{N\mathcal{Q}} \left[\eta(y)-\eta(x)\right]\,.
\end{equation}
For LD, and for any of the three models on regular graphs, $\mathcal{Q}$ is
constant.  Hence the summand in the expression on the right is antisymmetric
in $x$ and $y$ and $\langle \Delta\rho\rangle=0$.  As a consequence
$\langle\rho\rangle$ is conserved, so that the mutant fixation probability
equals the initial mutant density $\rho$.

To compute the exit probability on arbitrary networks for general models, we
generalize the notion of density by introducing the {\em degree-weighted
  moments}
\begin{equation}
  \label{omega}
\omega_{m} = \frac{1}{N\mu_{m}} \sum_x k_x^m \eta(x)= 
\frac{1}{\mu_{m}} \sum_k k^m n_k \rho_k\, ,
\end{equation}
where 
\begin{equation}
\label{VM:BG:rhok:def}
  \rho_k \equiv \frac{1}{N_k} \sideset{}{'}\sum_{x} \eta(x)\,, 
\end{equation}
is the density of $\mathbf{1}$s on the subset of nodes of degree $k$.  Here
the prime on the sum denotes the restriction that all nodes $x$ have fixed
degree $k$.  Note that $\omega_0$ coincides with the density $\rho$.  To
obtain a conserved quantity, it is clear that the factor $\mathcal{Q}$ in the
denominator of the transition rate in \eqref{master} must be canceled out.
For the VM, we thus consider $\Delta\langle\omega_1\rangle$ and repeat the
calculation that led to Eq.~\eqref{drho-av} to obtain
\begin{equation}
\label{omega1}
\Delta\langle\omega_1\rangle = \sum_{x,y}\frac{A_{xy}}{Nk_x}k_x\left[\eta(y)-\eta(x)\right].
\end{equation}
Similarly, the IP, we consider $\Delta\langle\omega_{-1}\rangle$ and obtain
\begin{equation}
\label{omega-1}
\Delta\langle\omega_{-1}\rangle = \sum_{x,y}\frac{A_{xy}}{Nk_x k_y}
\left[\eta(y)-\eta(x)\right].
\end{equation}
Because of the $x$-$y$ antisymmetry of the summand in Eqs.~\eqref{omega1} and
\eqref{omega-1}, both sums vanish so that the conserved quantity in the three
models are:
\begin{equation}
\label{cons}
\begin{tabular}{ll}
 $\langle\omega_1\rangle$  & \qquad \qquad  VM,\\
 $\langle\omega_0\rangle \equiv \langle\rho\rangle$ & \qquad \qquad LD,\\
 $\langle\omega_{-1}\rangle$ &  \qquad\qquad IP.\\
\end{tabular}
\end{equation}

We can understand these conservation laws intuitively.  For the VM, although
nodes are selected uniformly, there are relatively more low-degree nodes that
are neighbors of high-degree nodes (see Fig.~\ref{model}).  Thus low-degree
nodes change their state more often than high-degree nodes.  Weighting each
node by its degree compensates this disparity and leads to the conserved
quantity $\omega_1$.  Thus the mean density $\rho$ is not conserved, as first
pointed out in Ref.~\cite{SEM} for the VM on heterogeneous graphs.
In contrast, For the IP low-degree nodes change their state less often than
high degree nodes, and this disparity may be compensated by weighting each
node by its inverse degree.  This leads to the conserved quantity
$\omega_{-1}$.

\begin{figure}[ht]
\begin{center}
\includegraphics[width=0.125\textwidth]{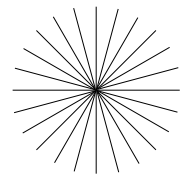}
\end{center}
\caption{A star graph.}
\label{star}
\end{figure}

Since the initial value of the conserved quantity equals its value in the
final unanimous state, we obtain, for the exit probability
\begin{equation}
\label{exit-cons}
\begin{tabular}{ll}
 $\mathcal{E}_\mathbf{1}(\omega_1)=\omega_1$ ~~~~~&  VM,\\
 $\mathcal{E}_\mathbf{1}(\omega_0)=\omega_0 \equiv \rho$ ~~~~~& LD \& degree
 regular graphs,\\
 $\mathcal{E}_\mathbf{1}(\omega_{-1})=\omega_{-1}$ ~~~~~& IP,\\
\end{tabular}
\end{equation}
An instructive example is the extreme case of a star graph, where $N$ nodes
are connected only to a single central hub (Fig.~\ref{star}).  For the VM, if
the hub is in state $\mathbf{1}$ and all other nodes are in state
$\mathbf{0}$, then Eq.~\eqref{exit-cons} predicts that the probability of
reaching $\mathbf{1}$ consensus is 1/2\,!  Thus a single individual with a
macroscopic number of neighbors largely determines the final state.
Conversely, for the IP, this same initial state reaches $\mathbf{1}$
consensus with probability that is $\mathcal{O}(1/N^2)$.  We will discuss
this dramatic disparity between VM and IP dynamics in detail in
Sec.~\ref{sec:evo}.

\section{VOTER MODEL ON THE COMPLETE GRAPH}
\label{sec:vm-cg}

As a preliminary for degree-heterogeneous graphs, we discuss the well-known
dynamics of the VM on the complete graph \cite{L,E,R01}, where each node is
connected to every other node.  Because the complete graph is degree regular,
the VM, IP, and LD are all equivalent and we treat the system in the
framework of the VM.  Let $\rho(t)$ be the density of voters in state
$\mathbf{1}$.  In each update event $\rho\to\rho\pm \delta\rho$, with
$\delta\rho=1/N$, corresponding to the respective state changes
$\mathbf{0}\to\mathbf{1}$ or $\mathbf{1}\to\mathbf{0}$.  The probabilities
for these events are
\begin{equation}
\label{VM:KG:jump} 
\begin{split}
  \mathbf{R}(\rho)\equiv \mathbf{P}[\rho\rightarrow\rho+\delta \rho] = 
(1-\rho)\rho \\
   \mathbf{L}(\rho)\equiv \mathbf{P}[\rho\rightarrow\rho -\delta \rho] = 
\rho(1-\rho).
\end{split}
\end{equation}
Here $\mathbf{R}$ and $\mathbf{L}$ denote raising and lowering operators that
give the transition probabilities from $\rho$ to $\rho\pm \delta\rho$,
respectively.

Let $c(\rho,t)$ be the probability that the density of $\mathbf{1}$'s is
$\rho$ at time $t$.  After one update event, this density evolves according
to
\begin{eqnarray}
\label{KG:occevo}
  c (\rho,t+\delta t)&=& \mathbf{R}(\rho-\delta\rho)c(\rho-\delta\rho,t)\nonumber\\
&~&+ \mathbf{L}(\rho+\delta\rho)c(\rho+\delta\rho,t)\nonumber\\
&~&+ [1-\mathbf{R}(\rho)-\mathbf{L}(\rho)]c(\rho,t). 
\end{eqnarray}
Here $\delta t=1/N$ and the first two terms on the right account for the
inflow to the state with density $\rho$ in an update and the last term
accounts for outflow.  Expanding Eq.~(\ref{KG:occevo}) to second order in
$\delta\rho$ gives the forward Kolmogorov or Fokker-Planck equation
\cite{VK},
\begin{equation}
  \frac{\partial c(\rho,t)}{\partial t} = \frac{\partial}{\partial\rho}
 [v(\rho)c(\rho,t)]+\frac{\partial^2}{\partial \rho^2} [D (\rho)c(\rho,t)], 
  \label{KG:KFP}
\end{equation}
where
\label{KG:drift}
\[v(\rho)\equiv \frac{\delta\rho}{\delta t} [\mathbf{R}(\rho)-\mathbf{L}(\rho)], 
\]
is the drift term (called the selection term in biology \cite{K83}) that is
caused by the bias in the transition probabilities, and
\begin{equation*}
\label{KG:diffusion}
  D(\rho)\equiv \frac{1}{2} \frac{\delta\rho^2}{\delta t} [\mathbf{R}(\rho)
+ \mathbf{L}(\rho)], 
\end{equation*}
is the diffusion term (paradoxically called the random drift term in biology)
that quantifies the stochastic noise in the kinetics.  On the complete graph,
the selection term is zero and the Fokker-Planck equation \eqref{KG:KFP}
becomes
\begin{equation}
\label{VM:KG:FP}
  \frac{\partial  c(\rho,t)}{\partial t} = \frac{1}{N}
  \frac{\partial^2}{\partial \rho^2} [\rho(1-\rho)c(\rho,t)]. 
\end{equation}

In a similar fashion, the equation for the exit probability with initial
density $\rho$ is
\begin{eqnarray}
\label{E}
  \mathcal{E}_\mathbf{1}(\rho)&=&\mathbf{R}(\rho)\mathcal{E}_\mathbf{1}(\rho\!+\!\delta\rho)
 +\mathbf{L}(\rho)\mathcal{E}_\mathbf{1}(\rho\!-\!\delta\rho) \nonumber\\
  &~&+ [1-\mathbf{R}(\rho)-\mathbf{L}(\rho)] \mathcal{E}_\mathbf{1}(\rho).
\end{eqnarray}
This equation expresses $\mathcal{E}_\mathbf{1}(\rho)$ as the probability of
making a transition to $\rho\pm \delta\rho$ or $\rho$, respectively, times
the exit probability from this intermediate point \cite{R01}.  Expanding
Eq.~\eqref{E} to second order in $\delta \rho$ gives the backward Kolmogorov
equation for the exit probability
\begin{equation}
  v(\rho)\frac{d \mathcal{E}_\mathbf{1}(\rho)}{d\rho}+D(\rho)\frac{d^2\mathcal{E}_\mathbf{1}(\rho)}{d \rho^2} \equiv
  \mathbf{B}\, \mathcal{E}_\mathbf{1}= 0,
\end{equation}
with $\mathbf{B}$ the {\em generator} of the backward Kolmogorov equation.
For the boundary conditions $\mathcal{E}_\mathbf{1}(\rho\!=\!0)=0$ and
$\mathcal{E}_\mathbf{1}(\rho\!=\!1)=1$, the solution is simply
$\mathcal{E}_\mathbf{1}(\rho)=\rho$.  This result reproduces
Eq.~\eqref{exit-cons} that was obtained by magnetization conservation.

In analogy with Eq.~\eqref{E}, the average time to reach consensus,
$T(\rho)$, as a function of the initial density $\rho$, obeys the backward
Kolmogorov equation \cite{R01}
\begin{eqnarray}
\label{T}
  T (\rho)&=& \delta t +  \mathbf{R}(\rho)T(\rho\!+\!\delta\rho)
 +\mathbf{L}(\rho) T(\rho\!-\!\delta\rho) \nonumber\\
  &~&+ [1-\mathbf{R}(\rho)-\mathbf{L}(\rho)] T(\rho).
\end{eqnarray}
This equation expresses the average consensus time as the time for a single
step plus the average time to reach consensus after taking this step.  The
three terms account for the transitions $\rho \rightarrow \rho\pm\delta\rho$
or $\rho \rightarrow \rho$, respectively, and the factor $\delta t$ in each
term accounts for the time elapsed in a single update.  Expanding
Eq.~\eqref{T} to second order in $\delta \rho$ gives the backward Kolmogorov
equation for the average consensus time
\begin{equation}
\label{T-back}
 v(\rho)\frac{d T(\rho)}{d\rho}+D(\rho)\frac{d^2 T(\rho)}{d \rho^2} \equiv
 \mathbf{B}\, T= -1.
\end{equation}

Using the transition probabilities in Eqs.~(\ref{VM:KG:jump}) and setting
$\delta t = \delta \rho = 1/N$, the above equation reduces to
\begin{equation}
 \frac{\rho(1-\rho)}{N}\,\frac{d^2 T(\rho)}{d\rho^2}=-1.
  \label{VM:KG:contime:eq}
\end{equation}
For the boundary conditions $T(0)= T(1)= 0$, the solution is
\begin{equation}
\label{VM:KG:contime:sol}
  T(\rho)= N\left[(1-\rho)\ln\frac{1}{1-\rho}+\rho\ln \frac{1}{\rho}\right] ,
\end{equation}
which is symmetric about $\rho=1/2$.  Important special cases are
$T(1/2)=N\ln 2$, corresponding to starting with equal densities of voters of
each opinion, and $T(1/N)\approx \ln N$, corresponding to starting with a
single mutant.

In addition to the time to reach either type of consensus, consider the {\em
  fixation times}, namely, the conditional times to reach $\mathbf{1}$
consensus, defined as $T_\mathbf{1}(\rho)$, or $\mathbf{0}$ consensus,
$T_\mathbf{0}(\rho)$, as a function $\rho$.  We obtain these times by
extending the backward Kolmogorov approach \cite{R01} to account for the
conditioning on type of consensus.  The conditional fixation times satisfy
\begin{equation}
\label{T-fix}
\mathbf{B}\left[\mathcal{E}_\mathbf{0}T_\mathbf{0}\right]=-\mathcal{E}_\mathbf{0}\qquad
\mathbf{B}\left[\mathcal{E}_\mathbf{1}T_\mathbf{1}\right]= -\mathcal{E}_\mathbf{1}, 
\end{equation}
with $\mathcal{E}_\mathbf{0}= 1- \mathcal{E}_\mathbf{1}$ and $\mathbf{B}$ is
the generator in Eq.~\eqref{T-back}, subject to absorbing boundaries for both
$\mathcal{E}_\mathbf{0}T_\mathbf{0}$ and
$\mathcal{E}_\mathbf{1}T_\mathbf{1}$.  The solution to Eq.~\eqref{T-fix} is
\begin{equation}
\label{condtime1}
T_\mathbf{0}(\rho)= N \frac{\rho}{1-\rho} \ln\frac{1}{\rho}\quad
T_\mathbf{1}(\rho)= N \frac{1-\rho}{\rho} \ln\frac{1}{1-\rho}~ .
\end{equation}
These fixation times then satisfy the sum rule
$\mathcal{E}_\mathbf{1}(\rho)T_\mathbf{1}(\rho)+
\mathcal{E}_\mathbf{0}(\rho))T_\mathbf{0}(\rho)=T(\rho)$ that reflects the fact that
the consensus time is the suitably weighted average of the fixation times.
One important limit is the initial state of a single $\mathbf{1}$ mutant, for
which the fixation time to all $\mathbf{1}$s is $T_\mathbf{1}(1/N)\approx N$.
In contrast, the consensus time from this same starting state is much
smaller, $T(1/N)\approx \ln N$, because the system can exit via the nearby
boundary at $\rho=0$.

\section{VOTER MODEL ON COMPLEX NETWORKS}
\label{vmk}

\subsection{The Complete Bipartite Graph\label{VM:Sec:BG}}

\begin{figure}[ht]
\begin{center}
\includegraphics[width=0.3\textwidth]{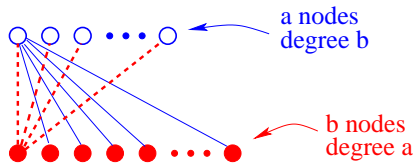}
\end{center}
\caption{(Color online) The complete bipartite graph $K_{a, b}$.}
\label{Kab}
\end{figure}

To understand how degree dispersity affects VM dynamics, we first study a
simple degree-heterogeneous network, the complete bipartite graph $K_{a,b}$,
in which $a+b$ nodes are partitioned into two subgraphs of size $a$ and $b$
(Fig.~\ref{Kab}).  Each node in the $\mathbf{a}$ subgraph is connected only
to all nodes in $\mathbf{b}$, and vice versa.  Thus $\mathbf{a}$ nodes all
have degree $b$, while $\mathbf{b}$ nodes all have degree $a$.

Consider the voter model on this graph.  Let $N_{\mathbf{a},\mathbf{b}}$ be
the respective number of voters in state $\mathbf{1}$ on each subgraph and
let $\rho_\mathbf{a} = N_\mathbf{a}/a$, $\rho_\mathbf{b} = N_\mathbf{b}/b$ be
the respective subgraph densities.  In an update, these numbers change
according to transition probabilities,
\begin{equation}
  \label{VM:BG:jump} 
\begin{split}
  \mathbf{R}_\mathbf{a} \equiv \mathbf{P}[\rho_\mathbf{a},\rho_\mathbf{b} 
\rightarrow \rho_\mathbf{a}^+,\rho_\mathbf{b}] 
= \frac{a}{a+b}\,\,\rho_\mathbf{b} (1-\rho_\mathbf{a}), \\
  \mathbf{L}_\mathbf{a} \equiv \mathbf{P}[\rho_\mathbf{a},\rho_\mathbf{b}
 \rightarrow \rho_\mathbf{a}^-, \rho_\mathbf{b}] 
= \frac{a}{a+b}\,\, \rho_\mathbf{a} (1-\rho_\mathbf{b}), 
\end{split}
\end{equation}
with $\rho_\mathbf{a}^\pm=\rho_\mathbf{a}\pm a^{-1}$.  Here
$\mathbf{R}_\mathbf{a}$ is the probability to increase the number of
$\mathbf{1}$s in subgraph $\mathbf{a}$ by 1, for which we need to first
choose a $\mathbf{0}$ in subgraph $\mathbf{a}$ that then interacts with a
$\mathbf{1}$ in subgraph $\mathbf{b}$.  Similarly, $\mathbf{L}_\mathbf{a}$
gives the corresponding the probability for reducing the number of
$\mathbf{1}$s in $\mathbf{a}$.  Analogous definitions hold for
$\mathbf{R}_\mathbf{b}$ and $\mathbf{L}_\mathbf{b}$ by interchanging
$a\leftrightarrow b$.

{}From these transition probabilities, the rate equations for average subgraph
densities are $\frac{d\rho_{\mathbf{a},\mathbf{b}}}{dt}=
\rho_{\mathbf{b},\mathbf{a}}-\rho_{\mathbf{a},\mathbf{b}}$ \cite{SR}, with
solution
\begin{eqnarray}
\label{rhoab}
\rho_{\mathbf{a},\mathbf{b}}(t)\!=\!  \frac{\rho_\mathbf{a}(0)\!+\! \rho_\mathbf{b}(0)}{2}\!+\!
\frac{  [\rho_{\mathbf{a},\mathbf{b}}(0)\!-\!\rho_{\mathbf{b},\mathbf{a}}(0)] e^{-2 t}}{2}~ . 
\end{eqnarray}
Thus the subgraph densities are driven to the common value
$(\rho_\mathbf{a}(0)+ \rho_\mathbf{b}(0))/2$ in a time of the order of one.
As a result, the density of $\mathbf{1}$s in the entire graph, which evolves as
\begin{eqnarray*}
\frac{d\rho}{dt} = \frac{1}{a+b} \left(a\frac{d\rho_\mathbf{a}}{dt}
+ b\frac{d\rho_\mathbf{b}}{dt}\right)=  \frac{b-a}{a+b} (\rho_\mathbf{a}-\rho_\mathbf{b}),
\end{eqnarray*}
becomes conserved in the long-time limit.  Thus there is a two-time scale
approach to consensus: at early times, there is a non-zero bias that quickly
drives the system to equal subgraph densities
$\rho_\mathbf{a}=\rho_\mathbf{b}$; subsequently, diffusive fluctuations drive
the eventual approach to consensus.

\begin{figure}[ht]
\begin{center}
\includegraphics[width=0.456\textwidth]{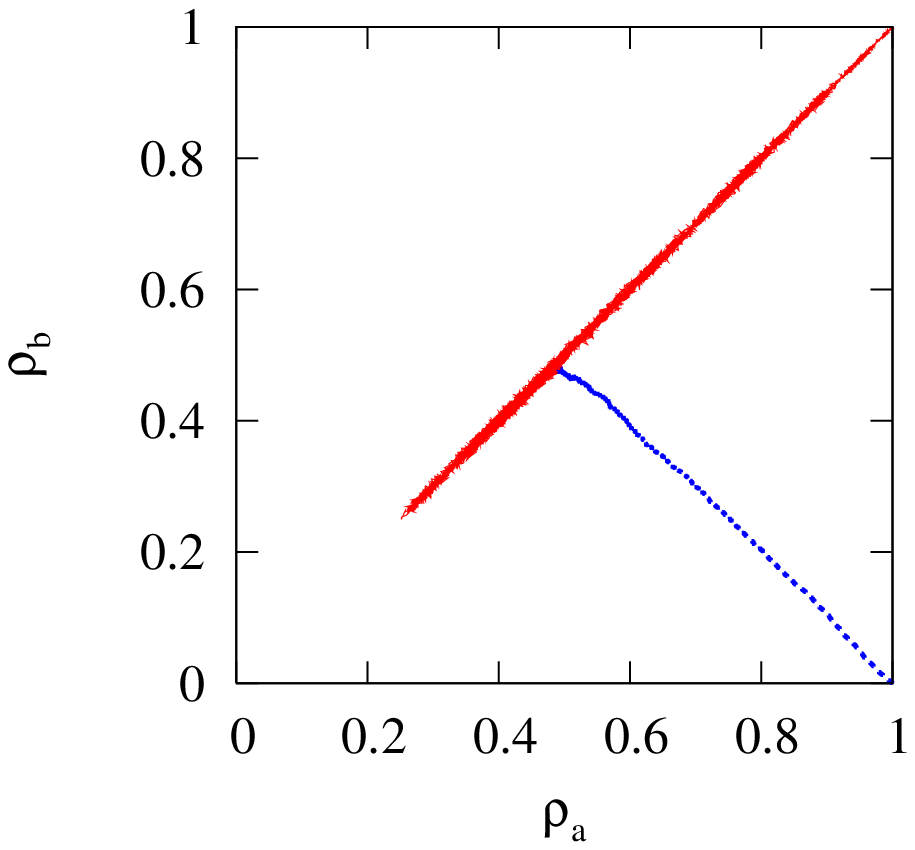}
\includegraphics[width= 0.456\textwidth]{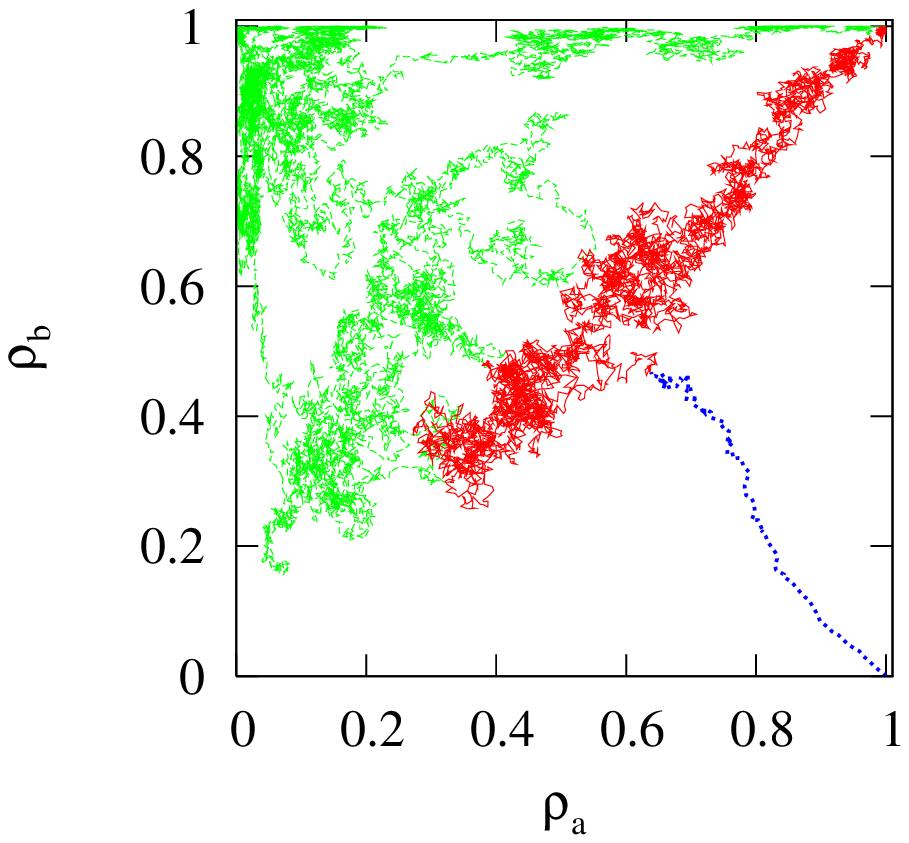}
\mbox{\hskip -0.45in \includegraphics*[width=0.425\textwidth]{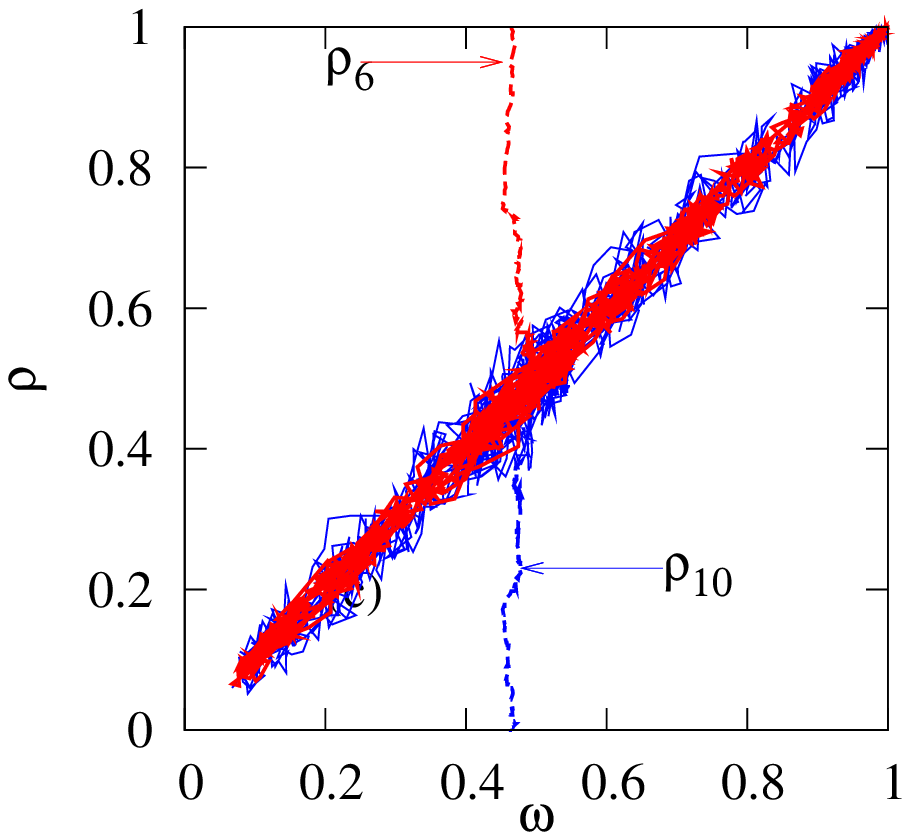}}
\end{center}
\caption{(Color online) Subgraph densities $\rho_\mathbf{b}(t)$ versus
  $\rho_\mathbf{a}(t)$ for single realizations of the voter model on (top to
  bottom): (a) a bipartite graph of $2\times 10^5$ nodes; (b) a 2-clique
  graph of $2\times 10^4$ nodes, with $C=1$ (upper trajectory) and $C=100$
  (lower trajectory); (c) a Molloy-Reed graph of $2\times 10^5$ nodes with
  degree distribution $n_k\sim k^{-2.5}$.  For the last example, the
  densities $\rho_6$ and $\rho_{10}$ are shown. }

\label{VM:BG:Fig:Evo}
\end{figure}

We can understand this behavior in a more fundamental way by studying the
probability that the graph has density of $\mathbf{1}$s equal to
$\rho_\mathbf{a}$ and $\rho_\mathbf{b}$ in each subgraph at time $t$,
$c(\rho_\mathbf{a},\rho_\mathbf{b},t)$.  This probability density evolves as
\begin{eqnarray}
\label{C}
  &&c(\rho_\mathbf{a},\rho_\mathbf{b},t+\delta t)= \nonumber \\
&~&~~~\mathbf{R}_\mathbf{a}(\rho_\mathbf{a}^-,\rho_\mathbf{b})\,
c(\rho_\mathbf{a}^-,\rho_\mathbf{b},t)
  +\mathbf{L}_\mathbf{a}(\rho_\mathbf{a}^+,\rho_\mathbf{b})\, 
c(\rho_\mathbf{a}^+,\rho_\mathbf{b},t) \nonumber \\
  &~&~~~+\mathbf{R}_\mathbf{b}(\rho_\mathbf{a},\rho_\mathbf{b}^-)\,
c(\rho_\mathbf{a},\rho_\mathbf{b}^-,t)
+ \mathbf{L}_\mathbf{b}(\rho_\mathbf{a},\rho_\mathbf{b}^+)\,
c(\rho_\mathbf{a},\rho_\mathbf{b}^+,t)  \nonumber \\
  &~&~~~~~+ \big[1-\mathbf{R}_\mathbf{a}(\rho_\mathbf{a},\rho_\mathbf{b})
-\mathbf{L}_\mathbf{a}(\rho_\mathbf{a},\rho_\mathbf{b})  \nonumber\\
&~&~~~~~~~-\mathbf{R}_\mathbf{b}(\rho_\mathbf{a},\rho_\mathbf{b})
-\mathbf{L}_\mathbf{b}(\rho_\mathbf{a},\rho_\mathbf{b})\big]\, 
c(\rho_\mathbf{a},\rho_\mathbf{b},t).
\end{eqnarray}
Expanding to second order in $a^{-1}$ and $b^{-1}$ gives the Fokker-Planck
equation,
\begin{eqnarray}
\label{VM:BG:FP:ver1} 
  \frac{\partial c}{\partial t} &\!\!=\!\!&-
\frac{1}{a\delta t} \frac{\partial}{\partial \rho_\mathbf{a}}
[(\mathbf{R}_\mathbf{a}-\mathbf{L}_\mathbf{a})\,c\,] 
-\frac{1}{b\delta t}\frac{\partial}{\partial\rho_\mathbf{b}}
[(\mathbf{R}_\mathbf{b}-\mathbf{L}_\mathbf{b})\,c\,]\nonumber\\
&~&+  \frac{1}{2a^2\delta t} \frac{\partial^2}{\partial\rho_\mathbf{a}^2}
[(\mathbf{R}_\mathbf{a}\!+\!\mathbf{L}_\mathbf{a})\,c\,] + \frac{1}{2b^2\delta t}
\frac{\partial^2}{\partial\rho_\mathbf{b}^2}
[(\mathbf{R}_\mathbf{b}\!+\!\mathbf{L}_\mathbf{b})\,c\,].\nonumber \\
\end{eqnarray}
Using $\delta t=1/(a+b)$, we identify the drift velocities for the two
subgraph densities as
\begin{eqnarray}
  \label{VM:BG:drift:sol}  
   v_\mathbf{a} &\equiv& \frac{a+b}{a} (\mathbf{R}_\mathbf{a}-\mathbf{L}_\mathbf{a})
= \rho_\mathbf{b}-\rho_\mathbf{a}, \nonumber\\
   v_\mathbf{b} &\equiv& \frac{a+b}{b} (\mathbf{R}_\mathbf{b}-\mathbf{L}_\mathbf{b})
= \rho_\mathbf{a}-\rho_\mathbf{b},
\end{eqnarray}
that again illustrates the early-time bias in the VM dynamics on the complete
bipartite graph.  This two-time-scale dynamics is illustrated in
Fig.~\ref{VM:BG:Fig:Evo}, where a bipartite network of size $a=b=10^5$ is
initialized with $(\rho_\mathbf{a}(0),\rho_\mathbf{b}(0))=(1,0)$.  The dotted
curve shows the convective transient that lasts for 3 time steps before the
subsequent diffusive approach to the final consensus after 9999.0 steps
(solid curve).  We define the end of the transient when the trajectory first
approaches to within the extent of diffusive fluctuations about the diagonal.

To determine the exit probability, we use the fact that $\omega\equiv
(\rho_\mathbf{a}+\rho_\mathbf{b})/2$ is conserved [see
Eq.~\eqref{exit-cons}], so that
\begin{equation}
\label{VM:BG:exit}
\mathcal{E}_\mathbf{1}(\rho_\mathbf{a},\rho_\mathbf{b}) = \omega(\rho_\mathbf{a},\rho_\mathbf{b})=
\frac{1}{2} (\rho_\mathbf{a}+\rho_\mathbf{b}). 
\end{equation}
Strikingly, when one subgraph contains only $\mathbf{0}$s and the other only
$\mathbf{1}$s, the probabilities of ending with all $\mathbf{1}$s is 1/2,
\textit{independent} of the subgraph size.  To determine the average time
$T(\rho_\mathbf{a},\rho_\mathbf{b})$ to reach consensus---either all
$\mathbf{1}$s or all $\mathbf{0}$s---as a function of $\rho_\mathbf{a}$ and
$\rho_\mathbf{b}$, we follow the same steps \cite{SR} as in Eqs.~\eqref{T}
and \eqref{T-back} to obtain the backward Kolmogorov equation
\begin{eqnarray}
\label{VM:BG:CT:eq:con}
  N \delta t &\!=\!& (\rho_\mathbf{a}-\rho_\mathbf{b})
\left(\frac{\partial T}{\partial\rho_\mathbf{a}}-
\frac{\partial T}{\partial\rho_\mathbf{b}}\right)\nonumber\\
&~&-\frac{1}{2}(\rho_\mathbf{a}\!+\!\rho_\mathbf{b}\!-\!2 \rho_\mathbf{a}\rho_\mathbf{b})
\left(\frac{1}{a} \frac{\partial^2 T}{\partial \rho_\mathbf{a}^2}+\frac{1}{b} 
\frac{\partial^2T}{\partial\rho_\mathbf{b}^2} \right).~~~~~~ 
\end{eqnarray}
The first term on the right accounts for the bias that drives the system to
equal subgraph densities while the second term accounts for the diffusion
that ultimately drives the system to consensus.

Since the subgraph densities $\rho_\mathbf{a}$ and $\rho_\mathbf{b}$
asymptotically approach each other and $\omega=
(\rho_\mathbf{a}+\rho_\mathbf{b})/2$, we replace $\rho_\mathbf{a}$ and
$\rho_\mathbf{b}$ by $\omega$ and also $\frac{\partial~~~}{\partial
  \rho_{\mathbf{a},\mathbf{b}}}$ by
$\frac{1}{2}\frac{\partial}{\partial\omega}$ in Eq.~\eqref{VM:BG:CT:eq:con}
to yield
\begin{equation}
\label{VM:BG:CT:omega:eq}
  \omega (1-\omega)\frac{\partial^2T}{\partial\omega^2} =-\frac{4ab}{a+b}.
\end{equation}
For the boundary conditions $T(0)=T(1)=0$, the solution is [compare with
Eq.~(\ref{VM:KG:contime:sol})],
\begin{equation}
\label{VM:BG:CT:sol}
  T(\omega)= \frac{4 a b}{a+b} \left[(1-\omega)\ln\frac{1}{1-\omega}
 +\omega\ln\frac{1}{\omega}\right] .
\end{equation}
The consensus time has a similar form to that of the complete graph, but with
an {\em effective} population size $N_{\mathrm{eff}} \equiv 4ab/(a+b)$.  If
both components of a bipartite graph have a similar size, $a, b\approx N/2$,
then $N_{\rm eff}\approx N$.  However, if the two components are disparate,
{\it e.g.}, for $a\sim \mathcal{O}(1)$ and $b\approx N$ then $T\sim
\mathcal{O}(1)$\,!  Thus one highly-connected node strongly facilitates
reaching consensus.

\subsection{The Two-Clique Graph}

\begin{figure}[ht]
\begin{center}
\includegraphics[width=0.325\textwidth]{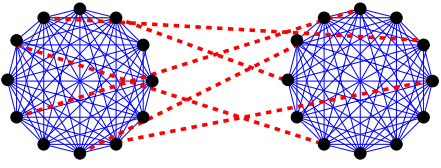}
\end{center}
\caption{(Color online) The two-clique graph with $N=12$ and $C=1/2$.}
\label{2c}
\end{figure}

Another instructive example that reveals the two-time-scale route to
consensus is the two-clique graph (Fig.~\ref{2c}).  This graph consists of
two separate complete graphs of $N$ nodes, $\mathbf{a}$ and $\mathbf{b}$,
with each node in one clique also possessing $C$ random cross links to the
nodes in the other clique.  As we now show, the extent of cross-linking
strongly affects how the system reaches consensus.

For the two-clique network, the respective probabilities that the clique
density $\rho_{\mathbf{a}}$ increases or decreases are:
\begin{equation}
\begin{split}
  &\mathbf{R}_{\mathbf{a}}= \frac{1}{2} (1-\rho_\mathbf{a})
\left[ \frac{N}{N+C}\,\rho_\mathbf{a} +  \frac{C}{N+C}\,\rho_\mathbf{b}\right]\\
  &\mathbf{L}_{\mathbf{a}}= \frac{1}{2} \rho_\mathbf{a}
\left[ \frac{N}{N+C}\,(1-\rho_\mathbf{a}) +  
\frac{C}{N+C}\,(1-\rho_\mathbf{b})\right]\,,
\end{split}
\end{equation}
with analogous expressions for the rates on clique $\mathbf{b}$.  For
$\mathbf{R}_{\mathbf{a}}$, the leading factor $\frac{1}{2}$ is the
probability of choosing clique $\mathbf{a}$ and $(1-\rho_\mathbf{a})$ is the
probability of choosing a $\mathbf{0}$ in clique $\mathbf{a}$.  The density
of $\mathbf{1}s$ in clique $\mathbf{a}$ increases if we choose a $\mathbf{0}$
in clique $\mathbf{a}$ (factor $\frac{N}{N+C}\,\rho_\mathbf{a}$) or in clique
$\mathbf{b}$ (factor $\frac{C}{N+C}\,\rho_\mathbf{b}$).  Similar explanations
apply for $\mathbf{L}_{\mathbf{a}}$ and for the operators on clique
$\mathbf{b}$.

The drift velocity for the density $\rho_{\mathbf{a}}$ is then
\begin{eqnarray}
\label{two-clique-drift}
  v_{\mathbf{a}} &=&\frac{\delta\rho_{\mathbf{a}}}{\delta t}\,
(\mathbf{R}_{\mathbf{a}}-\mathbf{L}_{\mathbf{a}})
  =\frac{C}{N+C}\,\left(\rho_{\mathbf{b}}-\rho_{\mathbf{a}}\right).
\end{eqnarray}
When $C$ is $\mathcal{O}(1)$, the drift term is $\mathcal{O}(1/N)$ and is
therefore comparable to the diffusive term,
\begin{equation}
  D_{\mathbf{a}}=\frac{1}{2} \frac{(\delta\rho_{\mathbf{a}})^2}{\delta t} 
  (\mathbf{R}_{\mathbf{a}}+\mathbf{L}_{\mathbf{a}})
  \propto \frac{(1/N)^2}{1/N}\,\frac{N}{N+C}~,
\end{equation}
which remains $\mathcal{O}(1/N)$ for all densities.  The clique densities
$\rho_{\mathbf{a}}$ and $\rho_{\mathbf{b}}$ thus evolve diffusively and
independently until consensus which is reached.  This behavior is illustrated
in the upper dashed (green online) trajectory of Fig.~\ref{VM:BG:Fig:Evo}(b)
for a single realization of a 2-clique graph with $a = b = 10^4$ nodes in
each clique and with a single crosslink per node for the initial condition
$(\rho_\mathbf{a}(0),\rho_\mathbf{b}(0))=(0,1)$.

Conversely, for $C>\mathcal{O}(1)$ the drift is dominant and the clique
densities are driven to a common value that subsequently evolves diffusively
along the $\rho_{\mathbf{a}}=\rho_{\mathbf{b}}$ diagonal until consensus is
reached as illustrated in the lower trajectory in
Fig.~\ref{VM:BG:Fig:Evo}(b).  Shown is the trajectory of a single realization
when there are 100 crosslinks per node for the initial condition
$(\rho_\mathbf{a}(0),\rho_\mathbf{b}(0))=(1,0)$.  The dotted (blue online)
line shows the initial transient of 100 time steps and the solid (red online)
line shows the subsequent diffusive approach to consensus at $\rho_a = \rho_b
= 1$ in 6437 time steps.

\subsection{Heterogeneous-Degree Networks}
\label{subsec:HDN}

We now study the VM on networks with arbitrary degree distributions.
Following Eq.~\eqref{master}, the fundamental transition probabilities for
increasing and decreasing the density of voters of type $\mathbf{1}$ on nodes
of fixed degree $k$ are:
\begin{equation}
  \label{VM:HG:jump} 
\begin{split}
  \mathbf{R}_k[\{\rho_k\}] &\!\!\equiv\!\! \mathbf{P}[\rho_k\rightarrow\rho_k^+]
  = \frac{1}{N} \sideset{}{'}\sum_{x,y}\!\frac{1}{k_{x}}\, A_{xy}\,\Phi(y,x)\\
  \mathbf{L}_k [\{\rho_k\}] &\!\!\equiv\!\! \mathbf{P}[\rho_k\rightarrow\rho_k^-]
  = \frac{1}{N} \sideset{}{'}\sum_{x,y}\!\frac{1}{k_{x}}\, A_{xy}\, \Phi(x,y),
\end{split}
\end{equation}
where $\rho_k^{\pm}=\rho_k\pm N_k^{-1}$, and the prime on the sums again
denote the restriction to nodes $x$ with fixed degree $k$.  In
Eq.~\eqref{VM:HG:jump} the densities associated with nodes of degrees $k'\ne
k$ are unaltered.

We now make the approximation that the degrees of neighboring nodes are
uncorrelated, such as in the Molloy-Reed (MR) network \cite{MR}.  Then we may
replace the elements of the adjacency matrix elements by their expected
values to give
\begin{equation}
\label{Amf}
  A_{x y} \to \langle A_{x y} \rangle = \frac{k_{x} k_{y}}{\mu_1 N} . 
\end{equation}
This relation expresses the fact that in the absence of degree correlations,
the probability that nodes $x, y$ are connected is proportional to $k_x k_y$,
and the proportionality constant in \eqref{Amf} is determined by using the
fact that the average node degree is just $\mu_1=\frac{1}{N}\sum_{x,y}
A_{xy}$ Substituting the mean-field assumption \eqref{Amf} for $A_{xy}$ in
Eq.~\eqref{VM:HG:jump}, and using Eqs.~\eqref{degreedis}, \eqref{omega}, and
\eqref{VM:BG:rhok:def}, the transition probabilities simplify to
\begin{equation}
\label{VM:RL}
  \mathbf{R}_k =  n_k\omega(1-\rho_k),\qquad  \mathbf{L}_k = n_k\rho_k(1-\omega).
\end{equation}

Similar to Eq.~(\ref{T}), the recursion formula for the mean consensus time,
starting with initial densities $\{\rho_k\}$, is
\begin{eqnarray}
  \label{VM:KBP}
 T(\{\rho_k\})&\!\!=\!\!& \delta t+\sum_k\mathbf{R}_k[\{\rho_k\}] T(\rho_k^+)
 +\sum_k \mathbf{L}_k[\{\rho_k\}]  T(\rho_k^-)\nonumber \\
 &~&+\Big[1-\sum_k(\mathbf{R}_k[\{\rho_k\}]+\mathbf{L}_k[\{\rho_k\}])\Big] T(\{\rho_k\}).
\nonumber\\
\end{eqnarray}
Expanding \eqref{VM:KBP} to second order in $N_k^{-1}$ gives the
backward Kolmogorov equation for the consensus time
\begin{equation}
\label{VM:Kol-t}
  \sum_k  v_k \frac{\partial T}{\partial\rho_k} +
  \sum_k  D_k \frac{\partial^2 T}{\partial\rho_k^2} =-1, 
\end{equation}
with degree-dependent velocity and diffusion coefficients
\begin{equation}
\label{VM:vD}
\begin{split}
  v_k &\equiv \frac{\delta \rho_k}{\delta t}(\mathbf{R}_k-\mathbf{L}_k)= \omega-\rho_k,\\
  D_k & \equiv \frac{(\delta\rho_k)^2}{2\delta t} (\mathbf{R}_k 
+ \mathbf{L}_k)= \frac{\omega+\rho_k-2 \omega \rho_k}{2Nn_k} ~.
\end{split}  
\end{equation}

To simplify the backward equation \eqref{VM:Kol-t} for the consensus time, we
start with the forward Kolmogorov equation for the probability distribution
\begin{equation}
\begin{split}
c(\{\rho_k\},t\!+\!\delta t) &\!=\! \sum_k\mathbf{F}_k[\rho_k-\delta_k] c(\rho_k-\delta_k,t)\\
&\!+\! \sum_k\mathbf{B}_k[\rho_k+\delta_k] c(\rho_k+\delta_k,t)\\
&\!+\!\Big[1\!-\!\sum_k(\mathbf{F}_k[\{\rho_k\}]\!+\! 
\mathbf{B}_k[\{\rho_k\}])\Big] c(\{\rho_k\},t),
  \end{split}
  \label{VM:KFP}
\end{equation}
and expand to second order to obtain the Fokker-Planck equation
\begin{equation}
  \partial_t c(\{\rho_k\},t) = - \sum_k\frac{\partial}{\partial\rho_k} v_k c + \sum_k
  \frac{\partial^2}{\partial\rho_k^2} D_k c . 
  \label{VM:HG:KFP}
\end{equation}
Now we compute the mean value of the average density $\rho_k$ with respect to
the above probability distribution to obtain the time dependence
\begin{equation}
  \frac{d \langle\rho_k\rangle}{d t} = \langle \omega\rangle-\langle\rho_k \rangle.
  \label{VM:HG:density:ExpChange}
\end{equation}
whose solution is, using the conservation of $\omega$,
\begin{equation}
  \langle\rho_k(t)\rangle  = \omega(0)-[\omega(0)-\rho_k(0)] e^{-t} .
  \label{VM:HG:rho:ExpSol}
\end{equation}
Thus after a time scale that is of the order of one, all the $\rho_k$
approach the common value of the conserved quantity $\omega$ and the drift
velocity in \eqref{VM:vD} vanishes.  This dynamics is illustrated in
Fig.~\ref{VM:BG:Fig:Evo}(c) where the trajectory of a single VM realization
is shown for a Molloy-Reed network of $2 \times 10^5$ nodes, with degree
distribution $n_k\sim k^{-2.5}$.  The initial state is
$(\rho_{k>\mu_1}(0),\rho_{k\leq\mu_1}(0)) =(0,1)$.  Shown are $\rho_{6}(t)$
(degree less than $\mu_1=8$) and $\rho_{10}(t)$ (degree greater than $\mu_1$)
versus $\omega$.  The initial transient, which lasts 1.78 time steps, is
shown dotted, while consensus occurs after 1742 time steps.

As a result of this rapid approach to the locus $\rho_k=\omega$ for each $k$, we
may drop the drift term in Eq.~\eqref{VM:Kol-t} and also convert to
derivatives with respect to $\omega$ by using
\begin{equation*}
\frac{\partial T}{\partial\rho_k}=\frac{\partial T}{\partial\omega}
\frac{\partial\omega}{\partial\rho_k} = \frac{k n_k}{\mu_1}\frac{\partial T}{\partial\omega},
\end{equation*}
to reduce \eqref{VM:Kol-t} to
\begin{equation}
\label{VM:HG:BPE}
  \frac{1}{N} \sum_k \left(\frac{k^2}{\mu_1^2} n_k\right)\omega(1-\omega)\,\,
\frac{\partial^2 T}{\partial \omega^2} =-1. 
\end{equation}
We now define the effective population size by
\begin{equation}
  \label{VM:HG:Neff}
  N_{\mathrm{eff}} = N\,\frac{\mu_1^2}{\mu_2},
\end{equation}
and comparing Eq.~\eqref{VM:HG:BPE} with \eqref{VM:KG:contime:eq}, the 
consensus time is
\begin{equation}
\label{TVM}
  T_N(\omega)= N_{\rm eff} \left[(1-\omega)\ln\frac{1}{1-\omega} 
+ \omega \ln \frac{1}{\omega}\right] ~. 
\end{equation}

To compute $N_{\rm eff}$ for a network of $N$ nodes with a power-law degree
distribution, $n_k \sim k^{-\nu}$, we first determine the maximum degree in
the network.  We obtain this quantity from the extremal criterion \cite{KR02},
$\int_{k_{\textrm{max}}} k^{-\nu}\, dk = \frac{1}{N}$.  This condition gives
$k_{\textrm{max}} \sim N^{1/(\nu-1)}$, which holds for all $\nu\geq 2$, while
for $\nu<2$, the maximum degree remains of the order of $N$.  With this upper
cutoff, the $m^{\textrm{th}}$ moment of the degree distribution is
\begin{eqnarray}
\label{moments}
\mu_m &\!\!\sim\!\!&
\int^{k_{\textrm{max}}} k^m\, n_k \, dk \nonumber \\ &\!\!\sim\!\!&
\begin{cases} N^{(m\!+\!1\!-\!\nu)/(\nu\!-\!1)}& m\!+\!1\!-\!\nu>0,~ \nu\geq2\\
 N^{(m\!+\!1\!-\!\nu)}& m\!+\!1\!-\!\nu>0,~\nu<2\\
\ln N & m\!+\!1\!-\!\nu=0\\
\mathcal{O}(1)& m\!+\!1\!-\!\nu<0.
\end{cases}
\end{eqnarray}
Using these results, the asymptotic behaviors of $N_{\rm eff}$ and $T_N$ are
\begin{equation}
\label{VM:HG:SF:CT}
 T_N \propto N_{\rm eff} \sim
\begin{cases}
N  & \nu>3,\\
N/\ln N & \nu=3,\\
N^{2(\nu-2)/(\nu-1)} & 2<\nu<3,\\
(\ln N)^2 & \nu=2,\\
{\cal O}(1)& \nu<2.
\end{cases}
\end{equation}

\begin{figure}[ht]
\begin{center}
\includegraphics[width=0.475\textwidth]{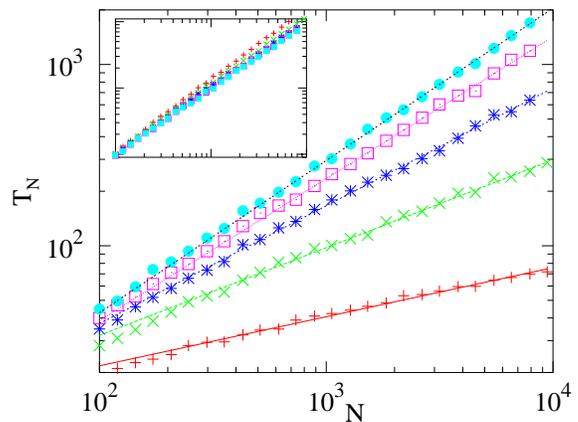}
\end{center}
\caption{(Color online) Consensus time $T_N$ versus $N$ on the Molloy-Reed
  network with degree distribution $n_k = k^{-\nu}$ for $\nu = 2.1$ $(+)$,
  $2.3$ $(\times)$, $2.5$ $(\ast)$, $2.7$ $(\circ)$ and $2.9$ $(\bullet)$.
  Each data point is based on 100 realizations of the graph and 10
  realizations of the voter model on each graph.  The lines represent the
  theoretical prediction of Eq.~\eqref{VM:HG:SF:CT}.  The inset shows the
  same data plotted in the scaled form $\mu_2T_N/\mu_1^2$ versus $N$.}
\label{TCU}
\end{figure}

The main feature of Eq.~\eqref{VM:HG:SF:CT} is that consensus is achieved
quickly for the VM, that is, $T_N\ll N$ for all $\nu< 3$ \cite{SR}.  This
consensus time is also much faster than the corresponding behavior on regular
lattices in $d$ spatial dimensions \cite{L,K92}: $T_N \propto N^2$ for $d=1$,
$T_N\propto N\ln N$ for $d=2$, and $T_N\propto N$ for $ d>2$.  We tested our
prediction \eqref{VM:HG:SF:CT} by simulations of the voter model on both the
MR network (Fig.~\ref{TCU}) and also a network that is grown by the the
redirection algorithm \cite{KR01}.  This method generates a network with
shifted linear attachment rate; that is, the probability of attaching to a
node of degree $k$ is given by $k+\lambda$.  This growth rule leads to a
power-law degree distribution network with tunable exponent $\nu=3+\lambda$.
For a network that is generated by the redirection algorithm, the $N$
dependence of $T_N$ is essentially the same as that for the MR network (see
Fig.~3 of Ref.~\cite{SR}), and both these numerical results are in good
agreement with our theory.

One important feature of the network that is built by the redirection
algorithm is that degrees of neighboring nodes are correlated \cite{KR01}.
In spite of this correlation, the actual values of the consensus times for
the MR and the redirection networks are numerically within 15\% of each other.
Thus evidently degree correlations have a secondary role in determining the
consensus time; the broadness of the degree distribution is much more
important.

\section{INVASION PROCESS}
\label{sec:IP}

We now study the role of degree heterogeneity on IP dynamics, a question that
was first studied by Castellano \cite{C05}.  In analogy with our approach for
the voter model, we replace the adjacency matrix elements by their average
values, as in Eq.~\eqref{Amf}, so that the transition probability of
Eq.~\eqref{master} becomes
\begin{equation}
\mathbf{P}[\eta\rightarrow\eta_{x}]=\frac{k_{x}}{\mu_1 N}[\eta(x)(1-\rho)+ (1-\eta(x))\rho].
\end{equation}
{}From this expression, and following exactly the same steps that led to
Eq.~\eqref{VM:RL}, the transition probabilities for nodes of a fixed given
degree are:
\begin{eqnarray}
 \label{VM:HG:IP:jump1} 
  \mathbf{R}_k = n_k \frac{k}{\mu_1} \rho (1-\rho_k)   \qquad
  \mathbf{L}_k   = n_k \frac{k}{\mu_1} \rho_k (1-\rho),
\end{eqnarray}
where again $\rho_k^\pm=\rho\pm N_k^{-1}$.  The corresponding $k$-dependent
drift velocity and diffusion coefficient are then:
\begin{equation}
\label{IP:HG:drifts}
\begin{split}
v_k &= \frac{\delta\rho_k}{\delta t}(\mathbf{R}_k-\mathbf{L}_k)=\frac{k}{\mu_1}(\rho-\rho_k),\\
D_k & = \frac{(\delta\rho_k)^2}{2\delta t} (\mathbf{R}_k 
+ \mathbf{L}_k)= \frac{k(\rho+\rho_k-2 \rho \rho_k)}{2Nn_k\mu_{-1}} 
\end{split}  
\end{equation}
{}From the drift velocity, the time dependence of the average density $\rho_k$
is determined from 
\begin{equation}
  \label{IP:HG:density:ExpChange}
  \frac{d \langle\rho_k\rangle}{d t} = \frac{k}{\mu_1} \langle( \rho-\rho_k)\rangle.
\end{equation}
which shows that the densities $\rho_k$ again approach the common value
$\rho$ after a time scale of $\mathcal{O}(1)$.  Once this concurrence
happens, we may asymptotically replace $\rho$, and concomitantly all the
$\rho_k$, by $\omega_{-1}$ in the expressions \eqref{IP:HG:drifts} for $v_k$
and $D_k$.

With this simplification, we now follow exactly the same steps that led from
Eq.~\eqref{VM:KBP} to Eq.~\eqref{TVM} in the previous section to write
backward Kolmogorov equation for the consensus time:
\begin{equation}
\label{VM:HG:IP:KBP}
  \frac{1}{\mu_1\mu_{-1}N}\,\, \omega_{-1}\, (1-\omega_{-1})\frac{\partial^2
    T}{\partial\omega_{-1}} =-1.
\end{equation}
By comparing with Eqs.~\eqref{TVM} and \eqref{VM:HG:IP:KBP}, we deduce the
average consensus time
\begin{equation}
\label{TIP}
T_N(\omega_{-1})= N_{\rm eff} \left[(1-\omega_{-1})\ln\frac{1}{1-\omega_{-1}} 
  + \omega_{-1} \ln \frac{1}{\omega_{-1}}\right]
\end{equation}
with $N_{\rm eff}=N\mu_1\mu_{-1}$.  For graphs with power-law degree
distributions we use the moments written in Eq.~\eqref{moments} to obtain
\begin{equation}
\label{T:IP}
T_N \propto N_{\rm eff} \sim
\begin{cases}
N  & \nu>2,\\
N\ln N & \nu=2,\\
N^{3-\nu} & \nu<2.
\end{cases}
\end{equation}
Thus for all reasonable graphs, those with $\nu>2$, the consensus time for the
IP is strictly linear in $N$, in sharp distinction to much faster consensus
for the VM.

\section{DYNAMICS WITH SELECTION}
\label{sec:evo}

We now study the VM and IP when the two states $\mathbf{0}$ and $\mathbf{1}$
have different fitnesses.  We define resident $\mathbf{0}$s as having fitness
$f=1$, and mutant $\mathbf{1}$s having fitness $f=r$, which can, in
principle, be either smaller or larger than 1.  Our goal is to determine the
{\em fixation probability}, namely, the probability that a single fitter
mutant ({\it i.e.}, we consider only the case $r>1$) overspreads a population
under biased dynamics \cite{ARS06}.  Such a phenomenon provides a natural
description for epidemic propagation \cite{AM,PV01,BBPV}, the emergence of
fads \cite{W,GH,BHW}, social cooperation \cite{nowak, szabo,SP}, or the
invasion of an ecological niche by a new species \cite{LHN,nowak,WB}.

We define the update steps in the biased VM as:
\begin{enumerate}[(i)]
\itemsep -1ex
\item pick a voter with probability proportional to its inverse fitness
  ($f^{-1}/N\langle f^{-1}\rangle$);
\item the voter adopts the state of a random neighbor.
\end{enumerate}
Here $\langle f^{-1}\rangle$ is the mean inverse fitness of the population.
Thus a weaker voter is more likely to be picked and to be influenced by a
neighbor.  We can equivalently view the fitness as the inverse of a death
rate.
Similarly, the evolution steps in the biased IP are:
\begin{enumerate}[(i)]
\itemsep -1ex
\item pick an invader with probability proportional to its fitness
  ($f/N\langle f\rangle$);
\item the invader exports its state to a random neighbor.
\end{enumerate}
We denote by $\langle f\rangle$ the mean fitness of the population that we
can equivalently view as the number of offspring produced by an individual.
Thus a fitter mutant is more likely to spread its progeny. For both the
biased VM and biased IP, we focus on the weak selection limit, with $r=1+s$
and $s\ll 1$.

Thus far, we have studied the VM and IP in discrete time, where the time is
incremented by $1/N$ after each update event.  For the biased models, it
turns out to be simpler to treat continuous-time versions.  Thus for the
biased VM with continuous dynamics, each individual adopts the state of a
random neighbor at a rate proportional to its inverse fitness.  Similarly, in
the biased IP, each individual exports its progeny to a random neighbor at a
rate proportional to its fitness.  This rate-based update rule is equivalent
to a discrete-time update with the time increment chosen from an exponential
distribution, with mean value $1/N \langle f^{-1}\rangle$ for the voter model
and $1/N \langle f\rangle$ for the invasion process.  The discrete and
continuous models are then equivalent except for this overall factor in the
time scale.

\subsection{Biased Voter Model}

For the biased VM on a general network, the probability of changing the state
at node $x$ is, in close analogy with the unbiased case [compare with
Eq.~\eqref{master}],
\begin{equation}
  \label{masterbias}
  \textbf{P} [\eta \to \eta_{x}] = \sum_{y} \frac{A_{x y}}{N k_x \langle f^{-1}\rangle}
\!\! \left[\Phi(y,x)+  \frac{1}{r}\, \Phi(x,y)\right],
\end{equation}
Following the same approach as in Sec.~\ref{subsec:HDN}, the density $\rho_k$
of $\mathbf{1}$s at nodes of degree $k$ increases by $\delta\rho_k=1/N_k$
with probability $\mathbf{R}_k(\eta)$ and decreases by $1/N_k$ with
probability $\mathbf{L}_k(\eta)$ in a single update, where
\begin{eqnarray}
    \mathbf{R}_k(\eta)&=& \frac{1}{Nk\langle f^{-1}\rangle}\sideset{}{'}\sum_{x y}
  A_{x y} \Phi(y,x)\nonumber \\
    \mathbf{L}_k(\eta)&=&\frac{1-s}{Nk\langle f^{-1}\rangle}\sideset{}{'}\sum_{x y}
  A_{x y} \Phi(x,y)
  \label{VM:bVM:jump1} 
\end{eqnarray}
are the respective transition probabilities for
($\mathbf{0}\rightarrow\mathbf{1)}$ and ($\mathbf{1} \rightarrow \mathbf{0})$
in the weak selection limit $s\ll 1$.  The primes on the sums again denote
the restriction to nodes $x$ of degree $k$.  We seek the fixation probability
$\mathcal{E}_\mathbf{1}$ to the state consisting of all $\mathbf{1}$s as a
function of the initial densities of $\mathbf{1}$.  This probability obeys
the backward Kolmogorov equation $\mathbf{B}\, \mathcal{E}_\mathbf{1}=0$
\cite{VK,R01}, subject to the boundary conditions $
\mathcal{E}_\mathbf{1}(\rho=0)=0$ and $ \mathcal{E}_\mathbf{1}(\rho=1)=1$.  In
the diffusion approximation, the generator $\mathbf{B}$ of this equation may
be expressed as a sum of the changes in $\rho_k$ over all $k$,
\begin{equation}
  \mathbf{B} = \frac{1}{\delta t} \sum_k \left[\delta \rho_k (\mathbf{R}_k
  -\mathbf{L}_k)\frac{\partial}{\partial\rho_k}
  + \frac{(\delta \rho_k)^2}{2} 
(\mathbf{R}_k + \mathbf{L}_k)\frac{\partial^2}{\partial\rho_k^2} \right] \,,
  \label{VM:bVM:KBG}
\end{equation}
with $\delta t = 1/N\langle f^{-1}\rangle$.

On degree-regular graphs, the sums in Eq.~(\ref{VM:bVM:jump1}) include all
nodes and hence count the total number of active links.
For uncorrelated node degrees, the fraction of active links reduces to
$\rho(1-\rho)$, and the generator \eqref{VM:bVM:KBG} for $s\ll 1$ becomes
\begin{equation}
\label{VM:bVM:RG:KBG}
  \mathbf{B} \approx  \rho(1-\rho) \left[s \frac{\partial}{\partial \rho}+\frac{1}{N} 
  \frac{\partial^2}{\partial \rho^2} \right]\,,
\end{equation}
The drift and diffusion terms differ by a factor $\mathcal{O}(sN)$, so that
selection dominates when the population size $N$ is larger than
$\mathcal{O}(1/s)$, while diffusion, or random genetic drift, dominates
otherwise.  Since the probability of increasing the density of \textbf{1}s at
each update is $r$ times larger than the probability of decreasing this
density, the fixation process is the same as the absorption of a uniformly
biased random walk in a finite interval.  The fixation probability is thus
given by the well-known exact formula \cite{E}
\begin{equation}
  \mathcal{E}_\mathbf{1}(\rho)= \frac{1-r^{-N\rho}}{1-r^{-N}} .
  \label{bLDsoldis}
\end{equation}
In the small-selection limit $s\ll 1$, this result approaches the solution of
the backward Kolmogorov equation $\mathbf{B}\,\mathcal{E}_\mathbf{1} = 0$,
with generator $\mathbf{B}$ given in \eqref{VM:bVM:RG:KBG},
\begin{equation}
\label{bLDsolcont}
\mathcal{E}_\mathbf{1}(\rho)=  \frac{1-e^{-sN\rho}}{1-e^{-sN}}\equiv \mathcal{F}(sN,\rho)\,,
\end{equation}
where for later usage we introduce the notation $\mathcal{F}$ for the
fixation probability.

Now we return to degree-heterogeneous networks.  Here the conserved quantity
for unbiased dynamics is the average degree-weighted density $\omega$
[Eq.~(\ref{cons})].  This conservation law suggests that we study the
evolution of $\langle\omega\rangle$ when $r\ne 1$.  When node $x$ is updated,
$\omega$ changes by
\begin{equation}
\omega(\eta_{x})-\omega(\eta) = k_{x}(1-2\eta(x))/\mu_1 N,
\end{equation}
where $\eta_x$ again denotes the state of the system after the update.  Thus
$\langle\omega\rangle$ evolves as
\begin{eqnarray}
  \label{VM:bVM:omega:evolve} 
  \frac{\partial\langle\omega\rangle}{\partial t} &=& \frac{1}{\delta t} \sum_{x} 
  [\omega(\eta_{x})-\omega(\eta)] \mathbf{P}[\eta\rightarrow\eta_{x}]   \nonumber\\
  &=& \frac{s}{\mu_1 N} \sum_{x, y} A_{x y}\Phi(x,y),
\end{eqnarray}
for $s\ll1$.  Now we use the mean-field assumption \eqref{Amf} for $A_{xy}$
to reduce Eq.~(\ref{VM:bVM:jump1}) to
\begin{equation}
  \label{VM:bVM:jump2}
  \begin{split}
  \mathbf{R}_k (\eta)&= \frac{1}{\langle f^{-1}\rangle}\, n_k \omega(1-\rho_k),\\
  \mathbf{L}_k (\eta)&=  \frac{1-s}{\langle f^{-1}\rangle}\, n_k \rho_k(1-\omega),
  \end{split}
\end{equation}
and Eq.~(\ref{VM:bVM:omega:evolve}) to
\begin{equation}
\label{VM:bVM:MR:omega:evolve}
 \frac{\partial\langle\omega\rangle}{\partial t}=
s\langle\omega\rangle (1-\langle\omega\rangle), 
\end{equation}
whose solution is
\begin{equation}
  \langle\omega(t)\rangle= \left\{1-[1-\omega(0)^{-1}] e^{-s t}\right\}^{-1}\, .
  \label{VM:bVM:MR:omega:sol}
\end{equation}

Then the time evolution of $\langle\rho_k\rangle$ becomes
\begin{equation}
  \frac{\partial\langle\rho_k\rangle}{\partial t}\! =\! 
  \frac{\delta \rho_k (\mathbf{R}_k\!-\!\mathbf{L}_k)}{\delta t} 
  = \langle\omega\rangle-\langle\rho_k\rangle +
s (1-\langle\omega\rangle)\langle\rho_k\rangle .
  \label{VM:bVM:rhok:evolve}
\end{equation}

\begin{figure}[ht]
\begin{center}
\includegraphics[width=0.45\textwidth]{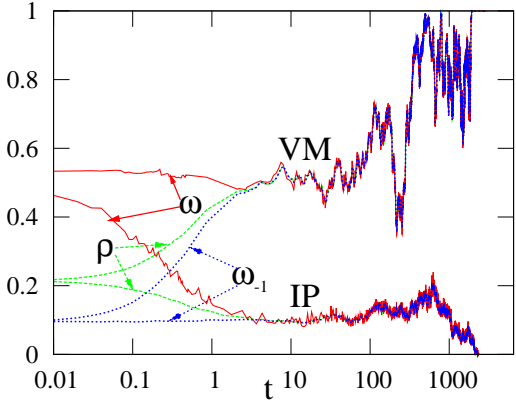}
\end{center}
\caption{(Color online) Illustration of the two-time-scale dynamics.  Moments
  of the \textbf{1} density in the biased VM and the biased IP on a
  Molloy-Reed network of $10^4$ nodes with a power-law degree distribution
  $n_k \sim k^{-\nu}$ and $\nu = 2.5$.  Nodes with degree larger than the
  mean degree are initialized to \textbf{1} while all other nodes are
  \textbf{0}.  Here $s = 8.13 \times 10^{-4}$ for the VM and $s = 6.6 \times
  10^{-5}$ for the IP.  These $s$ values were chosen to be $1/N_{\rm eff}$,
  with $N_{\rm eff}=N\mu_1^2/\mu_2$ for the VM and $N_{\rm
    eff}=N\mu_1\mu_{-1}$ for the IP.}
\label{VM:Fig:bVM:transient}
\end{figure}

To solve this equation we combine it with Eq.~(\ref{VM:bVM:MR:omega:evolve}) to
yield
\[\frac{\partial \langle\omega-\rho_k\rangle}{\partial t}
=-(1-s)\langle\omega-\rho_k\rangle
\langle 1-\omega\rangle,\] with solution
\begin{equation}
 \langle \rho_k(t)\rangle = \langle\omega(t)\rangle-e^{-t} [\omega (0)-\rho_k (0)] \{
  \omega (0)+ [1-\omega (0)] e^{-s t} \}\,. \label{VM:bVM:MR:rhok:sol}
\end{equation}
For small selective advantage ($s \ll 1$), this equation involves two
distinct time scales.  In a time of the order of 1, all the $\rho_k$ become
equal to $\omega$, the conserved quantity for the unbiased VM.  Subsequently,
the evolution of $\omega$ itself occurs on a longer time scale of order
$s^{-1} \gg 1$ and is driven by the bias (Fig.~\ref{VM:Fig:bVM:transient}).

\begin{figure}[ht]
\begin{center}
\includegraphics[width=0.425\textwidth]{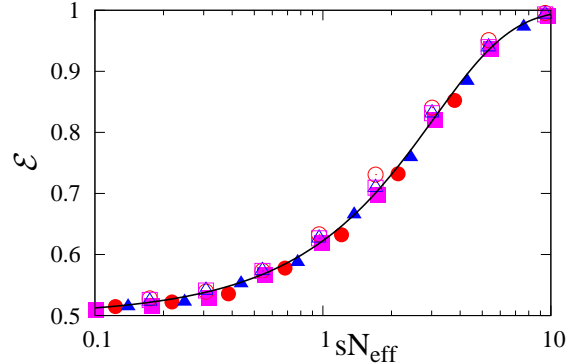}
\end{center}
\caption{(Color online) Scaling plot of fixation probabilities for VM and IP
  dynamics for a Molloy-Reed graph with degree distribution $n_k \sim
  k^{-\nu}$ and $\nu=2.5$, with $N=10^3$ and $\mu_1 = 8$.  The empty symbols
  correspond to IP dynamics with $s = 0.004$ ($\square$), $s = 0.008$
  ($\circ$) and $s = 0.016$ ($\triangle$); the filled symbols correspond to
  VM dynamics with $s = 0.01$, ($\blacksquare$), $s = 0.02$ ($\bullet$) and
  $s=0.08$ ($\blacktriangle$).  The smooth curve is the prediction of
  Eq.~\eqref{bLDsolcont}.}
\label{VM:fig:fixation:scaled}
\end{figure}

We now determine the fixation probability by replacing the $\rho_k$ by
$\omega$ in the transition probabilities $\mathbf{R}$ and $\mathbf{L}$ in
Eqs.~(\ref{VM:bVM:jump2}), and the derivative $\frac{\partial}{\partial
  \rho_k}$ by $\frac{kn_k}{\mu_1}\frac{\partial}{\partial \omega}$.  Then the
generator (\ref{VM:bVM:KBG}) becomes, in the limit $s\ll 1$,
\begin{eqnarray}
  \mathbf{B} &\approx & 
 \omega (1-\omega)\left[s \frac{\partial}{\partial\omega}+\frac{\mu_2}{\mu_1^2 N}
 \frac{\partial^2}{\partial\omega^2} \right],
  \label{VM:bVM:KBG:omega} 
\end{eqnarray}
which is the same as the generator for degree-regular graphs in
Eq.~(\ref{VM:bVM:RG:KBG}) with the population size $N$ replaced by $N_{\rm
  eff}=N\mu_1^2/\mu_2$.  Consequently, the solution to $\mathbf{B}\,
\mathcal{E}_\mathbf{1}= 0$ is simply $\mathcal{E}_\mathbf{1}(\omega)=
\mathcal{F}(sN_{\rm eff},\omega)$, with $\mathcal{F}$ defined in
Eq.~\eqref{bLDsolcont}, that is in excellent agreement with our simulation
data (Fig.~\ref{VM:fig:fixation:scaled}).

\begin{figure}[ht]
\begin{center}
\includegraphics[width=0.45\textwidth]{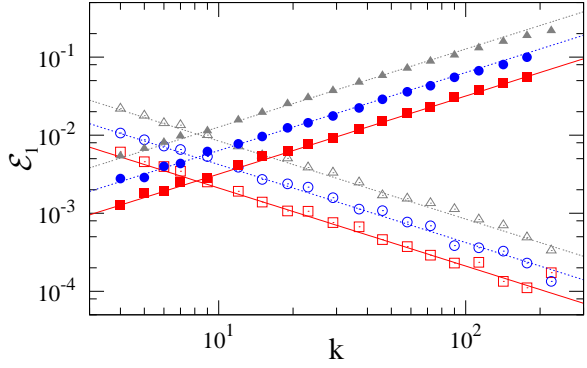}
\end{center}
\caption{(Color online) Fixation probability of a single mutant initially at
  a node of degree $k$ on a Molloy-Reed network with $n_k \sim k^{-\nu}$ and
  $\nu = 2.5$, with $N = 10^3$ and $\mu_1 = 8$. The empty symbols correspond
  to IP dynamics with $s = 0.004$ ($\square$), $s = 0.008$ ($\circ$) and $s =
  0.016$ ($\triangle$); the filled symbols correspond to VM dynamics with $s
  = 0.01$, ($\blacksquare$), $s = 0.02$ ($\bullet$) and $s = 0.08$
  ($\blacktriangle$). The solid lines, with slopes $+ 1$ and $- 1$,
  correspond to the second of Eqs.~(\ref{VM:bVM:fix:k:sol}) and
  (\ref{VM:bIP:fix:k:sol}).}
  \label{fix-k}
 \end{figure}

When a single mutant is initially located at a node of degree $k$, then
$\omega = k/N\mu_1$.  Substituting this value into $\mathcal{F}(sN_{\rm
  eff},\omega)$, we obtain the general result that the fixation probability
starting with a single mutant on a node of degree $k$ is proportional to $k$;
$\mathcal{E}_\mathbf{1}\propto k$ for {\em all\/} $s\ll 1$
(Fig.~\ref{fix-k}).  More precisely, $\mathcal{E}_\mathbf{1} \equiv
\mathcal{F}(sN_{\rm eff},\omega=k/N\mu_1)$ has two distinct limiting
behaviors:
\begin{equation}
\label{VM:bVM:fix:k:sol} 
\mathcal{E}_\mathbf{1}  \to 
\begin{cases}
    {\displaystyle \frac{1}{N \mu_1}\,\,k} & \quad s \ll 1/N_{\mathrm{eff}} ;\\ \\
       {\displaystyle  \frac{s  \mu_1}{\mu_2}\,\,k} & \quad  1/N_{\mathrm{eff}} \ll s \ll 1.
\end{cases}
\end{equation} 
In the $s \ll 1/N_{\mathrm{eff}}$ limit, we recover the results given in
Eq.~\eqref{exit-cons} for unbiased evolution.  Notice also that the relative
influence of selection and random genetic drift is determined by the variable
combination $sN_{\textrm{eff}}$.  Because $N_{\rm eff}$ can be much less than
$N$, diffusion can be important for much larger populations compared to
degree-regular graphs.

\subsection{Biased Invasion Process}

In the complementary biased invasion process, each individual reproduces at a
rate proportional to its fitness.  Hence the transition probability is
\begin{equation}
  \label{VM:bIP:jump0}
  \textbf{P} [\eta \to \eta_{x}] =   \sum_{y} \frac{A_{x y}}{N k_y\langle f\rangle}
  \left[r\, \Phi(y,x)+ \Phi(x,y)\right]\,.
\end{equation}
For degree-uncorrelated graphs and weak selection $s\ll1$, the transition probabilities for increasing and
decreasing $\rho_k$ in a single update become
\begin{equation}
  \begin{split}
  \mathbf{R}_k (\eta)&=    \frac{(1+s)k}{\mu_1\langle f\rangle}~  n_k \rho (1-\rho_k),\\
  \mathbf{L}_k (\eta)&= \frac{k}{\mu_1\langle f\rangle}~ n_k \rho_k (1-\rho) .
  \end{split}
  \label{VM:bIP:jump1} 
\end{equation}
Following the same steps that led to Eq.~(\ref{VM:bVM:rhok:evolve}) and using
$\delta t=1/N\langle f\rangle$ for continuous-time dynamics, $\rho_k$ evolves
as
\begin{equation}
\frac{\partial\rho_k}{\partial t} = \frac{k}{\mu_1} \left[\rho-\rho_k+s \rho_k (1-\rho)\right],
\end{equation}
which implies that all the $\rho_k$ rapidly become equal, and, as a
consequence, all moments $\omega_m$ also become equal.  For the unbiased IP,
the conserved quantity is $\omega_{-1}$; thus for weak selection,
$\omega_{-1}$ is now the most slowly changing quantity, as illustrated in
Fig.~\ref{VM:Fig:bVM:transient}.  Hence we replace all $\rho_k$ by
$\omega_{-1}$, and transform all derivatives with respect to $\rho_k$ to
derivatives with respect to $\omega_{-1}$ in the generator \eqref{VM:bVM:KBG}
to obtain
\begin{equation}
  \mathbf{B} = \frac{\omega_{-1} (1-\omega_{-1})}{\mu_1 \mu_{-1}} 
\left[s \frac{\partial}{\partial \omega_{-1}}+\frac{1}{N} 
  \frac{\partial^2}{\partial \omega_{-1}^2} \right] \label{VM:bIP:KBG}
\end{equation}
from which, in close analogy with our previous analysis of the VM, the
fixation probability is $\mathcal{E}_\mathbf{1}(\omega_{-1})=
\mathcal{F}(sN,\omega_{-1})$, with $\mathcal{F}$ again defined in
Eq.~\eqref{bLDsolcont}.

As a basic corollary, consider the fixation probability when the initial
state consists of a single mutant at a node of degree $k$.  Substituting
$\omega_{-1}=1/kN\mu_{-1}$ into $\mathcal{F}$, we obtain the general results
that in the small selection limit $s\ll 1$ the fixation probability is {\em
  inversely proportional\/} to the node degree, $\mathcal{E}_1\sim 1/k$
(fig.~\ref{fix-k}.  The limiting behaviors of the fixation probability are
\begin{equation}
  \label{VM:bIP:fix:k:sol} 
  \mathcal{E}_\mathbf{1} = 
\begin{cases}
{\displaystyle \frac{1}{N\mu_{-1}}\,\,\frac{1}{k}} &\quad s\ll 1/N ;\\ \\
 {\displaystyle \frac{s}{\mu_{-1}}\,\,\frac{1}{k}}  &\quad 1/N\ll s\ll 1.
\end{cases} 
\end{equation}
In the $s \ll 1/N$ limit we again recover the results in the neutral case
\eqref{exit-cons}.

\section{DISCUSSION}

We developed a unified framework to investigate the dynamics of the voter
model (VM), the invasion process (IP), as well as link dynamics (LD) on
complex networks.  When the relative fitness of the two states, $\mathbf{0}$
and $\mathbf{1}$, are the same, the variable $q$ defined by
\begin{equation}
\label{q-cases}
 q \equiv \begin{cases}
   \omega& \textrm{~~~~~VM},\\
   \rho & \textrm{~~~~~on regular graphs},\\
   \omega_{-1}  & \textrm{~~~~~IP},
\end{cases}
\end{equation}
is conserved by the dynamics.  Because of this conservation law, the exit
probability $\mathcal{E}_\mathbf{1}$ to $\mathbf{1}$ consensus is simply
equal to this conserved variable.  This simple result has far-reaching
consequences on complex networks, where high-degree nodes can have a
disproportionately strong influence.

On complex networks, the evolution of the VM and the IP is governed by two
disparate time scales.  Initially there is a quick approach to a homogeneous
state in which the density of $\mathbf{1}$s on nodes of degree $k$ becomes
independent of $k$ for any initial condition.  Diffusive fluctuations then
drive the system to a final consensus state on a much slower time scale.  For
a network of $N$ nodes with a power-law degree distribution, $n_k\sim
k^{-\nu}$, the VM consensus time $T_N$ scales slower than linearly with $N$
for all $\nu<3$.  This result seems to hold {\em independent\/} of the extent
of degree correlations.  Thus when the degree distribution is sufficiently
broad, high degree nodes facilitate the approach to consensus.  In contrast,
in the IP, $T_N$ scales linearly with $N$ for power-law degree networks for
all $\nu>2$.

We also studied the VM and IP with the additional feature of selection, in
which individuals in state $\mathbf{1}$ are fitter than those in the
$\mathbf{0}$ state.  The populations in these two models can be characterized
by an effective size
\begin{equation}
N_{\mathrm{eff}} =
\begin{cases}
{\displaystyle \frac{N\mu_1^2}{\mu_2}} &\quad{\rm VM},\\ \\
N\mu_1\mu_{-1}\sim N & \quad {\rm IP\ (\&\ regular\ graphs)}.  
\end{cases}
\end{equation}
The equivalence $N_{\rm eff}\sim N$ applies for the physically accessible
case where the mean degree of a graph is finite.  {}From the fixation
probabilities given in the previous section, the fundamental parameter for
both models is $\alpha \equiv sN_\mathrm{eff}$, which is just the P\'eclet
number in the language of biased diffusion.  The interesting limit is $s\to
0$ and $N\to\infty$, such that $\alpha$ is constant.  Now the general form of
the backward Kolmogorov generator is
\begin{equation}
\label{B-gen}
  \mathbf{B} = \frac{1}{ N_\mathrm{eff}}\,\, q(1-q)
  \left[\alpha \frac{\partial}{\partial q}+ 
  \frac{\partial^2}{\partial q^2} \right] \,.
\end{equation}

An important characteristic of the dynamics is the probability that a single
$\mathbf{1}$ mutant overspreads an otherwise homogeneous population of
$\mathbf{0}$s.  This fixation probability satisfies
$\mathbf{B}\,\mathcal{E}_\mathbf{1}(q)=0$, with solution
\begin{equation}
\label{fixgen}
 \mathcal{E}_\mathbf{1}(q)= \frac{1-e^{-\alpha q}} {1-e^{-\alpha}}~,
\end{equation}
that depends on the scaling combination $sN{\rm eff}$ and is independent of
$s$ and $N_\mathrm{eff}$ separately.  Another important feature of fixation
is its dependence on the degree of the node at which a mutant first appears.
For the VM, the probability for a mutant on a node of degree $k$ to fixate is
proportional to $k$ [Eq.~\eqref{VM:bVM:fix:k:sol}], while in the
complementary IP, the fixation probability is proportional to $1/k$
[Eq.~\eqref{VM:bIP:fix:k:sol}].  The origin of this behavior is simple to
understand.  In the VM, a well-connected mutant is more likely to be asked
its opinion before the mutant queries one of its neighbors.  In the IP, a
mutant on a high-degree node is more likely to be invaded by a neighbor
before the mutant itself can invade.  Thus network heterogeneity leads to
effective evolutionary heterogeneity.

The above results also allow us to understand the fixation probability for
the case where a mutant spontaneously appears at random in the network.  For
the VM in the selection-dominated regime ($sN_{\rm eff} \gg 1$), averaging
Eq.~\eqref{VM:bVM:fix:k:sol} over all nodes gives a fixation probability that
is smaller by a factor $\mu_1^2/\mu_2\le1$ than that on regular graphs.  Thus
a heterogeneous graph is an inhospitable environment for a mutant with VM
dynamics.  The source of this inhospitability is that a
spontaneously-appearing mutant is likely to be on a low-degree node, and its
state is then quickly erased by interactions with higher-degree nodes.
Conversely, for the IP, averaging Eq.~\eqref{VM:bIP:fix:k:sol} over all nodes
gives a fixation probability that is independent of the node degree.

Finally, with our general formalism, we can determine the average consensus
time from the solution of the backward Kolmogorov equation
$\mathbf{B}\,T(q)=-1$.  For the unbiased VM and the unbiased IP, this
solution has the generic form:
\begin{equation}
\label{TIPq}
  T_N(q)=  N_{\rm eff} \left[(1-q)\ln\frac{1}{1-q} +q\ln\frac{1}{q}\right] ~,
\end{equation}
where $q$ is the conserved quantity given in Eq.~\eqref{q-cases}.  In the
biased case the solution of Eq.~\eqref{B-gen} can be written as an integral
but it is quite cumbersome \cite{E}. From the form of the generator in
Eq.~\eqref{B-gen}, however, it is clear that $T_N(q)= N_{\rm eff}\,
\tau(q,\alpha)$, and $\tau(q, \alpha)$ is a function that is non singular at
$\alpha=0$.  Hence, the dependence on effective population size
$N_\mathrm{eff}$ is not affected by the bias when $\alpha$ is kept constant.

\acknowledgments{We acknowledge financial support to the Program for
  Evolutionary Dynamics at Harvard University by Jeffrey Epstein and NIH
  grant R01GM078986 (TA) and NSF grant DMR0535503 (SR \& VS).}


\begin{thebibliography}{99}

\bibitem{CEAH} R. Cohen, K. Erez, D. ben-Avraham, and S. Havlin, Phys.\ Rev.\
  Lett.\ {\bf 85}, 4626 (2000).

\bibitem{PV} R. Pastor-Satorras and A. Vespignani, Phys.\ Rev.\ E {\bf 63}, 066117 (2001).

\bibitem{DGM} S. N. Dorogovtsev, A. V. Goltsev, and J. F. F. Mendes, Phys.\
  Rev.\ E {\bf 66}, 016104 (2002).

\bibitem{L} T. M. Liggett, {\em Interacting Particle Systems},
  (Springer-Verlag, Berlin, 2005).

\bibitem{K92} P. L. Krapivsky, Phys.\ Rev.\ A {\bf 45}, 1067 (1992).

\bibitem{C05} C. Castellano, AIP Conference Proceedings {\bf 779}, 114 2005).

\bibitem{M} P. A. P Moran, {\em The Statistical Processes of Evolutionary
    Theory} (Clarendon Press, Oxford, 1962).

\bibitem{nowak} M. A. Nowak, {\em Evolutionary Dynamics}, (Harvard Univ.\
  Press, Cambridge MA 2006).

\bibitem{WB} M. C. Whitlock and N. H. Barton, Genetics {\bf 146}, 427 (1997).

\bibitem{LHN} E. Lieberman, C. Hauert, and M. A. Nowak, Nature {\bf 433}, 312 (2005).

\bibitem{SEM} K. Suchecki, V. M. Eguiluz, and M. San Miguel, Europhys.\
  Lett. {\bf 69} 228 (2005).

\bibitem{E} W. Ewens, {\em Mathematical Population Genetics I. Theoretical
  Introduction}, (Springer-Verlag, Berlin, 2004).

\bibitem{R01} S. Redner, {\em A Guide to First-Passage Processes}, (Cambridge
  University Press, New York, 2001).

\bibitem{VK} N. G. van Kampen, {\em Stochastic Processes in Physics and
    Chemistry}, $2^{\rm nd}$ ed.\ (North-Holland, Amsterdam, 1997).

\bibitem{K83} M. Kimura, {\em The Neutral Theory of Molecular Evolution},
  (Cambridge University Press, Cambridge, 1983).

\bibitem{SR} V. Sood and S. Redner, Phys.\ Rev.\ Lett.\ {\bf 94}, 178701
  (2005).

\bibitem{MR} M. Molloy and B. Reed, Random Structures \& Algorithms {\bf 6},
  161 (1995).

\bibitem{KR02} P. L. Krapivsky and S. Redner, J. Phys.\ A {\bf 35}, 9517 (2002).

\bibitem{KR01} P. L. Krapivsky and S. Redner, Phys.\ Rev.\ E {\bf 63}, 066123 (2001).

\bibitem{ARS06} T. Antal, S. Redner, and V. Sood, Phys.\ Rev.\ Lett.\ {\bf
    96}, 188104 (2006)

\bibitem{AM} R. M. Anderson and R. M. May, {\em Infectious Diseases in
    Humans}, (Oxford University Press, Oxford, 1992).

\bibitem{PV01} R. Pastor-Satorras and A. Vespignani, Phys.\ Rev.\ Lett.\ {\bf 86}, 3200 (2001).

\bibitem{BBPV} M. Barthelemy, A. Barrat, R. Pastor-Satorras, and
  A. Vespignani, Phys.\ Rev.\ Lett.\ {\bf 92}, 178701 (2004).

\bibitem{W} D. J. Watts, Proc.\ Natl.\ Acad.\ Sci.\ {\bf 99}, 5766 (2002).

\bibitem{GH} A. Gronlund and P. Holme, Adv.\ Complex Systems {\bf 8}, 261 (2005).

\bibitem{BHW} J. Bendor, B. A. Huberman, and F. Wu, arXiv.org:physics/0509217.

\bibitem{szabo} G. Szab\'o, G. F\'ath, Physics Reports {\bf 446}, 97 (2007).

\bibitem{SP} F. C. Santos and J. M. Pacheco, Phys.\ Rev.\ Lett.\ {\bf 95},
  098104 (2005).


\end{thebibliography}
\end{document}